\documentclass[10pt]{article}
\usepackage{dcolumn}
\usepackage{latexsym}
\usepackage{graphicx}
\usepackage{pstricks}
\usepackage{epsfig}
\usepackage{color}

\input epsf
\usepackage{amsmath,amssymb}
\newcommand{\be}{\begin{equation}}
\newcommand{\ee}{\end{equation}}
\newcommand{\bea}{\begin{eqnarray}}
\newcommand{\eea}{\end{eqnarray}}

\newcommand{\lbl}[1]{\label{eq:#1}}
\newcommand{ \rf}[1]{(\ref{eq:#1})}

\newcommand{\lapprox}{%
\mathrel{%
\setbox0=\hbox{$<$}
\raise0.6ex\copy0\kern-\wd0
\lower0.65ex\hbox{$\sim$}
}}
\newcommand{\gapprox}{%
\mathrel{%
\setbox0=\hbox{$>$}
\raise0.6ex\copy0\kern-\wd0
\lower0.65ex\hbox{$\sim$}
}}

\begin{document}
\allowdisplaybreaks

\renewcommand{\thefootnote}{\fnsymbol{footnote}}

\begin{center}
{\Large\bf On the contribution from the light quarks to $H\to\gamma\gamma , \gamma Z$} 

\vspace{0.75cm}

{Marc Knecht}$^1$\footnote{knecht@cpt.univ-mrs.fr} and Baiyi Qian$^{2,3}$\footnote{baiyi.qian@ijclab.in2p3.fr} 

\indent 

{$^1${\small\it{Centre de Physique Th\'{e}orique, Aix-Marseille Univ., Universit\'e de Toulon, CNRS (UMR7332)\\
CNRS-Luminy Case 907, 13288 Marseille Cedex 9, France}} } 

\vspace{0.2cm}

{$^2${\small\it{Facult\'e des Sciences, Aix-Marseille Universit\'e, 163 Avenue de Luminy, Case 901, 13288 Marseille Cedex 9, France}} } 

\vspace{0.2cm}

{$^3${\small\it{Universit\'e Paris-Saclay, CNRS/IN2P3, IJCLab, 91405 Orsay, France}} }

\indent

\begin{abstract}
\noindent
This Letter addresses the contribution from the light $u , d , s$ quarks to the amplitudes 
for the decay modes of the Brout-Englert-Higgs scalar boson $H$ into two photons ($H\to\gamma\gamma$) 
or to a photon and a neutral weak gauge boson ($H\to\gamma Z$), taking into account 
the non-perturbative aspects of QCD. Contrary to a recent claim, the 
contribution from the light quarks does not vanish. Rather, it is shown that, in contrast to the perturbative 
evaluations usually considered in the literature, this contribution to the amplitudes starts with a term
that is linear, and not quadratic, in the masses of the light quarks, thus pointing toward a sizeable enhancement
of their contribution to $H \to \gamma \gamma$.
\end{abstract}

\end{center}

\renewcommand{\thefootnote}{\arabic{footnote}}
\setcounter{footnote}{0}

\indent 

\indent

\section{ }
\setcounter{equation}{0}

The amplitudes for the decay modes of the Brout-Englert-Higgs scalar boson $H$ into two photons ($H\to\gamma\gamma$)
or a photon and a neutral weak gauge boson ($H\to\gamma Z$) are forbidden at tree level in the standard model, 
and receive contributions only at the loop level,
see refs. \cite{Ellis:1975ap,Shifman:1979eb,Gavela:1981ri} and \cite{Cahn:1978nz,Bergstrom:1985hp,Phan:2021pcc,Aiko:2023nqj},
respectively, as well as ref. \cite{Djouadi:2005gi} for an overview.
It is thus interesting to probe them with high precision in the hope of disclosing evidence for the 
existence of physics beyond the standard model. At the level of an effective low-energy 
description, the corresponding amplitudes can be obtained from dimension-five contact terms,
\be
{\cal L}_{\rm eff} \supset - \frac{1}{4} C_{H\gamma\gamma} H F^{\mu\nu} F_{\mu\nu} - \frac{1}{2} C_{H\gamma Z} H F^{\mu\nu} Z_{\mu\nu}
\lbl{eff_couplings}
\ee
where $F_{\mu\nu}$ and $Z_{\mu\nu}$ are the field-strength tensors of the photon and neutral weak boson fields, respectively.
The corresponding $H-\gamma-\gamma$ and $H-\gamma-Z$ couplings, $C_{H\gamma\gamma}$  and $C_{H\gamma Z}$, receive contributions from the 
standard model as well as from possible new physics. In this Letter we will exclusively focus our attention on the former.

\indent 

\noindent
In the standard model, the scalar boson couples to the charged gauge bosons and to 
the charged fermions through their masses, 
\be
{\cal L}_{\rm int}^{\rm SM} =  g M_W H (W^+ \cdot W^-) - \frac{g}{2 M_W} H J_{(H)} - e A_\mu J^\mu_{(\gamma)} - \frac{g}{2 c_w} \, Z_\mu J_{(Z)}^\mu
 + \cdots , 
\ee
where (in the case of quarks, colour indices, being contracted to form colour singlets, are not shown explicitly)
\be
J_{(H)} = \sum_f m_f {\bar\psi}_f \psi_f , ~~~
J^\mu_{(\gamma)} = \sum_f e_f {\bar\psi}_f \gamma^\mu \psi_f , ~~~
J_{(Z)}^\mu = \sum_f {\bar\psi}_f \gamma^\mu ( {\rm v}_f - {\rm a}_f \gamma_5 ) \psi_f , 
\ee
denote the fermionic bilinear operators
coupling to the electroweak scalar boson $H$, to the photon $\gamma$, and to the neutral weak gauge boson $Z$, respectively.
In the latter case, one has
\be
{\rm v}_f = I^3_f - 2 s_{\rm w}^2 e_f , ~~~ {\rm a}_f = I^3_f   ,
\ee
with $s_{\rm w} = \sin\theta_{\rm w}$, $c_{\rm w} = \cos\theta_{\rm w}$, where $\theta_{\rm w}$ is the weak mixing angle,
whereas $e_f$ is the electric charge of the fermion $f$ in units of the positron charge $e$
and $I^3_f$ its weak-isospin projection.

\indent 

\noindent
In the absence of higher-order electroweak corrections, the contribution of the fermion $f$ to the corresponding amplitudes,
\bea
{\cal A}_{\gamma\gamma}^{(f)\lambda_1\lambda_2} \!\!\!\!&=&\!\!\!\! g e^2 \frac{m_f}{2 M_W} \,
\Gamma^{(f) \mu\nu}_{\gamma\gamma} (q_1 , q_2) \epsilon_\mu^{\lambda_1} (q_1)^* \epsilon_\nu^{\lambda_2} (q_2)^* \Big\vert_{q_1^2=q_2^2=0, q_3^2 = M_H^2}   ,
 \nonumber\\
\\
{\cal A}_{\gamma Z}^{(f)\lambda_1\lambda_2} \!\!\!\!&=&\!\!\!\! g^2 e \frac{m_f}{4 c_{\rm w} M_W} \,
\Gamma^{(f) \mu\nu}_{\gamma Z} (q_1 , q_2) \epsilon_\mu^{\lambda_1} (q_1)^* \epsilon_\nu^{\lambda_2} (q_2)^*
 \Big\vert_{q_1^2=0, q_2^2=M_Z^2, q_3^2 = M_H^2}    ,
\nonumber
\eea
involves the connected three-point functions
\be
\Gamma_{\gamma X}^{(f) \mu\nu} (q_1 , q_2) = \int d^4 x \, e^{i q_1 \cdot x} \int d^4 y \, e^{i q_2 \cdot y}
\langle 0 \vert T \{ [{\bar\psi}_f \psi_f] (0) J_{(\gamma)}^\mu (x) J_{(X)}^\nu (y) \} \vert 0 \rangle_0 . ~~~ X = \gamma , Z .
\lbl{vertex}
\ee
The subscript ``0'' in the vacuum expectation value under the double integral indicates that it has to be evaluated in free field theory for the leptons 
and in pure QCD for the quarks. In both cases invariance under charge conjugation applies, and the axial part of the current $J_{(Z)}^\mu$ does not contribute. 
Then these vertex functions satisfy the Ward identities
\be
\{ q_1^\mu \,;\, q_2^\nu \} \Gamma_{\gamma X}^{(f) \mu\nu} (q_1 , q_2) = \{ 0 \,;\, 0 \}, ~~~ X = \gamma , Z ,
\ee
and thus take, as a consequence of invariance under Lorentz and parity transformations, the form
\be
\Gamma^{(f) \mu\nu}_{\gamma X} (q_1 , q_2) = {\cal F}_{\gamma X}^{(f)} (q_1^2 , q_2^2 , q_3^2) P^{\mu\nu} (q_1 , q_2) +
{\cal G}_{\gamma X}^{(f)} (q_1^2 , q_2^2 , q_3^2) Q^{\mu\nu} (q_1 , q_2)   ,
\ee 
with $q_3 = q_1 + q_2$ and
\be
P^{\mu\nu} (q_1 , q_2) = q_1^\nu q_2^\mu - (q_1 \cdot q_2) \eta^{\mu\nu} , ~~~
Q^{\mu\nu} (q_1 , q_2) = q_1^2 q_2^\mu q_2^\nu + q_2^2 q_1^\mu q_1^\nu - (q_1 \cdot q_2) q_1^\mu q_2^\nu - q_1^2 q_2^2 \eta^{\mu\nu}  .
\ee
Notice that due to the structure of the tensor $Q^{\mu\nu} (q_1 , q_2)$, its contraction  with the polarization vector 
$\epsilon_\mu^{\lambda_1} (q_1)^*$ of the emitted photon vanishes after taking $q_1^2=0$. The invariant functions ${\cal G}^{(f)}_{\gamma\gamma}$ 
and ${\cal G}^{(f)}_{\gamma Z}$ will thus not contribute to the decay amplitudes and need not be considered further.
In this language, the contribution at one loop from a charged fermion to the effective $H-\gamma-\gamma$ and $H-\gamma-Z$ couplings is simply given by
\be
C_{H\gamma\gamma}^{(f)} \big\vert_{\rm 1\,loop} = - \frac{g e^2}{2} \frac{m_f}{M_W} \,e_f^2 {\cal F}_{\gamma \gamma}^{(f)} (0 , 0 , M_H^2) ,~~~
C_{H\gamma\gamma}^{(f)} \big\vert_{\rm 1\,loop} = - \frac{g^2 e}{4 c_{\rm w}} \frac{m_f}{M_W} \, e_f {\rm v}_f {\cal F}_{\gamma Z}^{(f)} (0 , M_Z^2 , M_H^2)  ,
\lbl{C_pert}
\ee
where
\be
{\cal F}_{\gamma \gamma}^{(f)} (0 , 0 , M_H^2) = 4 {\cal F}^{(f)} ( 0 , 0 , M_H^2)  , ~~~
{\cal F}_{\gamma Z}^{(f)} (0 , M_Z^2 , M_H^2) = 4 {\cal F}^{(f)} ( 0 , M_Z^2 , M_H^2)   ,
\ee
with
\bea \lbl{calF_pert}
{\cal F}^{(f)} ( 0 , M^2 , M_H^2) \!\!\!\!&=&\!\!\!\!  4 N_c^{(f)} \, \frac{m_f}{M_H^2- M^2}  
\bigg\{ \frac{1}{16 \pi^2} \, \frac{M^2}{M^2 - 4 m_f^2} \, \frac{M_H^2 - 8 m_f^2}{M_H^2 - 4 m_f^2}
+ \frac{M^2}{M_H^2-M^2} \left[ {\bar J}_{(f)} (M_H^2) - {\bar J}_{(f)} (M^2) \right]
\nonumber\\
&&\hspace{2.5cm}
- \, \frac{M_H^2 - M^2 - 4 m_f^2}{M_H^2 - M^2} \, \frac{M^2}{M^2 - 4 m_f^2}
\, {\bar J}_{(f)} (M^2) \left[ 1 - 4 \pi^2 {\bar J}_{(f)} (M^2) \right]
\nonumber\\
&&\hspace{2.5cm}
+ \, \frac{M_H^2 - M^2 - 4 m_f^2}{M_H^2 - M^2} \, \frac{M_H^2}{M_H^2 - 4 m_f^2}
\, {\bar J}_{(f)} (M_H^2) \left[ 1 - 4\pi^2 {\bar J}_{(f)} (M_H^2) \right]  \bigg\}    .
\eea
Here ${\bar J}_{(f)} (s)$ denotes the one-loop two-point Feynman integral subtracted at $s=0$,
\be
{\bar J}_{(f)} (s) = \frac{s}{16\pi^2} \int_{4 m_f^2}^{+\infty} \frac{dx}{x} \, \sqrt{1 - \frac{4 m_f^2}{x}} \, \frac{1}{x-s-i \epsilon}   ,
\ee
while $N_c^{(f)} = 1$ when $f$ is a charged lepton, and $N_c^{(f)}=3$ when $f$ is a quark.
Within the perturbative framework, the suppression actually goes 
like the square of the fermion masses. The second mass factor comes from the loop
integral itself, ${\cal F}^{(f)} ( 0 , q^2 , M_H^2) \propto m_f$, which would actually vanish for a massless fermion. 
Therefore, the by-far most important contributions to the amplitudes for the decays $H\to\gamma\gamma$
and $H \to \gamma Z$ are provided by the loops involving the charged gauge bosons and the top quark, followed by 
the loops involving the $b$ and $c$ quarks or the $\tau$ lepton. Due to this mass suppression,
the remaining charged fermions, the light quarks $u$, $d$, $s$ and leptons $\mu$, $e$, only 
lead to tiny contributions, which are usually neglected. 
These considerations on the dependence with respect to the fermion masses are relevant as long as a perturbative 
approach is appropriate. While this is the case for the charged leptons and can be justified for the heavy 
quarks, it is more questionable in the case of the light quark flavours $u$, $d$ and $s$,
for which non-perturbative effects, like the spontaneous breaking of three-flavour 
chiral symmetry, play an important role.
Moreover, the dependence on their masses makes the contribution from the light quarks ambiguous,
since no intrinsic physical definition of the mass is available in this case. Using 
for instance the running mass defined in a given renormalization scheme like ${\overline{\rm MS}}$
introduces a dependence on the renormalization scale and scheme, which can, eventually, only be 
dealt with, at least at the perturbative level, upon considering higher-order QCD corrections.
This, however, still eludes the discussion of the non-perturbative QCD effects in the case of the light quarks.
For a discussion of the order ${\cal O} (\alpha_s)$ corrections and references to the original literature
see the review \cite{Djouadi:2005gi}.

\indent 

\noindent
To the best of our knowledge, the issue of the non-perturbative treatment of the 
contribution from the light quarks to the amplitudes of the decay modes $H\to\gamma\gamma , \gamma Z$
has actually never been tackled seriously. Recently, it has even been claimed that 
at the non-perturbative level it vanishes identically \cite{Hernandez-Juarez:2025ees}. This strong statement, 
which actually led us to undertake the present study, sounds rather puzzling. As we will show explicitly, 
it is based on some misconception and happens to be wrong. The correct treatment of the non-perturbative
aspects of QCD shows that the suppression due to the light-quark masses is still there, but it
is dominantly linear in these masses. In the case of $H \to \gamma \gamma$ the enhancement due to 
non-perturbative effects can actually be quite substantial, while it seems to remain rather marginal for $H \to \gamma Z$.
Furthermore, it is to be expected that any scale or scheme ambiguity coming from the mass dependence should disappear
at the non-perturbative level.

\section{ }
\setcounter{equation}{0}

Restricting from now on our attention to the light quark flavours $u,d,s$, and discarding
any electroweak corrections, we need to evaluate the corresponding three-point functions \rf{vertex} in three-flavour QCD.
The light-quark components of the currents appearing in the expressions of the three-point functions are then given by
\be
J_{(H)}^{\rm s} = \frac{{\rm tr} ({\cal M})}{3} {\bar\psi} \psi , ~~~ J_{(H)}^{\rm ns} = {\bar\psi} \stackrel{_{\ -\!\!-}}{\cal M}  \psi , ~~~ 
J_{(\gamma)}^\mu = {\bar\psi} {\cal Q} \gamma^\mu \psi , ~~~ 
J_{(Z)}^\mu = {\bar\psi} {\cal Q}_Z \gamma^\mu \psi - 2 s_{\rm w}^2 J_{(\gamma)}^\mu = - \frac{1}{6} {\bar\psi} \gamma^\mu \psi + 
( 1 - 2 s_{\rm w}^2 ) {\bar\psi} {\cal Q} \gamma^\mu \psi , 
\ee
where we have introduced a convenient notation that allows to consider the three flavours simultaneously, with
\be 
\psi \equiv \left(\!\!
\begin{tabular}{c}
$u$ \\ $d$ \\ $c$
\end{tabular}
\!\!\right) , ~~~
{\cal M} = {\rm diag} (m_u , m_d , m_s) , ~~~  \stackrel{_{\ -\!\!-}}{\cal M} \equiv {\cal M} - \frac{1}{3} {\rm tr} ({\cal M})  , ~~~
{\cal Q} = {\rm diag} \left( \frac{2}{3} , - \frac{1}{3} , - \frac{1}{3}  \right)
 , ~~~ {\cal Q}_Z = {\rm diag} \left( \frac{1}{2} , - \frac{1}{2} , - \frac{1}{2}  \right)   .
 \lbl{psi_M_Q}
\ee
Since QCD in invariant under charge conjugation, we have right away discarded the axial part of the weak current $J^\mu_{(Z)}$,
displaying only its vector part. We have further decomposed this vector current into a flavour singlet part and a flavour 
nonsinglet (octet) part, where the latter happens to be proportional to the electromagnetic current $J^\mu_{(\gamma)}$.
We have also performed a similar decomposition on the scalar density $J_{(H)}$, with its favour-singlet part 
denoted by $J_{(H)}^{\rm s}$ and its flavour-nonsinglet part by $J_{(H)}^{\rm ns}$. This leaves us with four independent vertex 
functions to consider,
\be
\Gamma_{H_{\rm ns}\gamma\gamma}^{\mu\nu} (q_1 , q_2) , ~~~
\Gamma_{H_{\rm s}\gamma\gamma}^{\mu\nu} (q_1 , q_2) , ~~~
\Gamma_{H_{\rm ns}\gamma Z_{\rm s}}^{\mu\nu} (q_1 , q_2) , ~~~
\Gamma_{H_{\rm s}\gamma Z_{\rm s}}^{\mu\nu} (q_1 , q_2)   ,
\lbl{vertex_list}
\ee
in an obvious notation and with definitions analogous to eq. \rf{vertex}. The subscripts ``s'' and ``ns'' refer to the 
flavour-singlet and flavour-nonsinglet components, respectively, of $J_{(H)}$ or $J_{(Z)}^\mu$. 

\indent

\noindent
These three-point functions all satisfy the Ward identities 
\be
\{ q_1^\mu \, ; \, q_2^\nu \} \Gamma^{\mu\nu} (q_1 , q_2) = \{ 0 \, ; 0 \} , ~~~
\Gamma \in \{ \Gamma_{H_{\rm ns}\gamma\gamma} , \Gamma_{H_{\rm s}\gamma\gamma} , \Gamma_{H_{\rm ns}\gamma Z_{\rm s}} , \Gamma_{H_{\rm s}\gamma Z_{\rm s}} \}   .
\lbl{WIs}
\ee
An additional simplification will be made at this stage: after factorization of the quark masses contained in $J^{\rm s,ns}_{(H)}$, 
we evaluate the three-point functions in the chiral limit, $m_u, m_d, m_s \to 0$,
since we want to focus on effects that are linear in the light-quark masses.  
When combined  with invariance under the Lorentz group, parity, charge conjugation and the Ward identities \rf{WIs}, this last condition
implies that the flavour and Lorentz structure of these correlators takes the form
\bea\lbl{vertex_def}
\Gamma_{H_{\rm ns}\gamma\gamma}^{\mu\nu} (q_1 , q_2) \!\!\!\!&=&\!\!\!\!
4 {\rm tr} (\stackrel{_{\ -\!\!-}}{\cal M} \! {\cal Q} {\cal Q} ) \times  
\left[ {\cal F}_{\gamma\gamma} (q_1^2 , q_2^2, q_3^2) P^{\mu\nu} (q_1 , q_2) + 
{\cal G}_{\gamma\gamma} (q_1^2 , q_2^2, q_3^2) Q^{\mu\nu} (q_1 , q_2) \right] , 
\nonumber\\
\Gamma_{H_{\rm s}\gamma\gamma}^{\mu\nu} (q_1 , q_2) \!\!\!\!&=&\!\!\!\!
\frac{4}{3} \, {\rm tr} ({\cal M}) {\rm tr} ( {\cal Q} {\cal Q} ) \times  
\left[ {\tilde{\cal F}}_{\gamma\gamma} (q_1^2 , q_2^2, q_3^2) P^{\mu\nu} (q_1 , q_2) + 
{\tilde{\cal G}}_{\gamma\gamma} (q_1^2 , q_2^2, q_3^2) Q^{\mu\nu} (q_1 , q_2) \right] , 
\\
\Gamma_{H_{\rm ns}\gamma Z_{\rm s}}^{\mu\nu} (q_1 , q_2) \!\!\!\!&=&\!\!\!\!
- \frac{2}{3} {\rm tr} (\stackrel{_{\ -\!\!-}}{\cal M} \! {\cal Q} ) \, \times  
\left[ {\cal F}_{\gamma Z} (q_1^2 , q_2^2, q_3^2) P^{\mu\nu} (q_1 , q_2) + {\cal G}_{\gamma Z} (q_1^2 , q_2^2, q_3^2) Q^{\mu\nu} (q_1 , q_2) \right] .
\nonumber
\eea 
The overall normalization factors have been chosen for later convenience.
The fourth three-point function in eq. \rf{vertex_list}, $\Gamma_{H_{\rm s}\gamma Z_{\rm s}}^{\mu\nu}$, 
has dropped out from this list since, under the assumptions just stated, it would be proportional 
to ${\rm tr} ({\cal Q})$ and thus vanishes. Finally, as mentioned earlier, only the invariant 
functions ${\cal F}_{\gamma\gamma}$, ${\tilde{\cal F}}_{\gamma\gamma}$ and ${\cal F}_{\gamma Z}$
will contribute to the decay amplitudes.

\indent 

\noindent
Separating the contributions from each light-quark flavour, one has
\bea 
{\cal A}_{\gamma\gamma}^{(u)\lambda_1\lambda_2}  \!\!\!\!&=&\!\!\!\!  
- \frac{g e^2}{2} \frac{m_u}{M_W} \, [\epsilon^{(\lambda_1)} (q_1) \cdot  \epsilon^{(\lambda_2)} (q_2) ]^* \, \frac{M_H^2}{2} 
\times e_u^2  \left[
2 {\cal F}_{\gamma\gamma} (0 , 0 , M_H^2) + 2 {\tilde{\cal F}}_{\gamma\gamma} (0 , 0 , M_H^2)
\right]     ,
\nonumber\\ 
{\cal A}_{\gamma\gamma}^{(d)\lambda_1\lambda_2}  \!\!\!\!&=&\!\!\!\!  
- \frac{g e^2}{2} \frac{m_d}{M_W} \, [\epsilon^{(\lambda_1)} (q_1) \cdot  \epsilon^{(\lambda_2)} (q_2) ]^* \, \frac{M_H^2}{2}
\times e_d^2 \left[
- 4 {\cal F}_{\gamma\gamma} (0 , 0 , M_H^2) + 8 {\tilde{\cal F}}_{\gamma\gamma} (0 , 0 , M_H^2)
\right]     ,
\nonumber\\
{\cal A}_{\gamma\gamma}^{(s)\lambda_1\lambda_2} \!\!\!\!&=&\!\!\!\!  
- \frac{g e^2}{2} \frac{m_s}{M_W} \, [\epsilon^{(\lambda_1)} (q_1) \cdot  \epsilon^{(\lambda_2)} (q_2) ]^* \, \frac{M_H^2}{2}
\times e_s^2 \left[
- 4 {\cal F}_{\gamma\gamma} (0 , 0 , M_H^2) + 8 {\tilde{\cal F}}_{\gamma\gamma} (0 , 0 , M_H^2)
\right]    ,
\eea
and
\bea\lbl{amps_uds} 
{\cal A}_{\gamma Z}^{(u)\lambda_1\lambda_2} \!\!\!\!&=&\!\!\!\!   
- \frac{g^2 e}{4 c_{\rm w}} \frac{m_u}{M_W} [\epsilon^{(\lambda_1)} (q_1) \cdot \epsilon_{Z}^{(\lambda_2)} (q_2) ]^* 
\, \frac{M_H^2 - M_Z^2}{2}
\times e_u
\bigg[ - \frac{2}{3} {\cal F}_{\gamma Z} (0 , M_Z^2, M_H^2) + \frac{4}{3} (1-2s_{\rm w}^2) {\cal F}_{\gamma\gamma} (0 , M_Z^2 , M_H^2) 
\nonumber\\
&& \hspace{7.6cm}
+ \, \frac{4}{3} (1-2s_{\rm w}^2)  {\tilde{\cal F}}_{\gamma\gamma} (0 , M_Z^2 , M_H^2)   
\bigg]   ,
\nonumber\\  
{\cal A}_{\gamma Z}^{(d)\lambda_1\lambda_2} \!\!\!\!&=&\!\!\!\!   
- \frac{g^2 e}{4 c_{\rm w}} \frac{m_d}{M_W} [\epsilon^{(\lambda_1)} (q_1) \cdot \epsilon_{Z}^{(\lambda_2)} (q_2) ]^*  
\, \frac{M_H^2 - M_Z^2}{2}
\times e_d \bigg[ - \frac{2}{3} {\cal F}_{\gamma Z} (0 , M_Z^2, M_H^2) + \frac{4}{3} (1-2s_{\rm w}^2) {\cal F}_{\gamma\gamma} (0 , M_Z^2 , M_H^2) 
\nonumber\\
&& \hspace{7.6cm}
- \, \frac{8}{3}  (1-2s_{\rm w}^2)  {\tilde{\cal F}}_{\gamma\gamma} (0 , M_Z^2 , M_H^2)   
\bigg]   ,
\nonumber\\  
{\cal A}_{\gamma Z}^{(s)\lambda_1\lambda_2} \!\!\!\!&=&\!\!\!\!   
- \frac{g^2 e}{4 c_{\rm w}} \frac{m_s}{M_W} [\epsilon^{(\lambda_1)} (q_1) \cdot \epsilon_{Z}^{(\lambda_2)} (q_2) ]^* 
\, \frac{M_H^2 - M_Z^2}{2}
\times e_s \bigg[ - \frac{2}{3} {\cal F}_{\gamma Z} (0 , M_Z^2, M_H^2) + \frac{4}{3} (1-2s_{\rm w}^2) {\cal F}_{\gamma\gamma} (0 , M_Z^2 , M_H^2) 
\nonumber\\
&& \hspace{7.6cm}
- \, \frac{8}{3}  (1-2s_{\rm w}^2)  {\tilde{\cal F}}_{\gamma\gamma} (0 , M_Z^2 , M_H^2)   
\bigg]  .
\eea
Since the functions between the straight brackets are evaluated in the chiral limit and, as we will 
show soon, do not vanish in this limit, we can already conclude that the contributions of the light 
quarks to the decay amplitudes will be linear in their masses, with a non-vanishing coefficient. 
In the sequel, we will address some non-perturbative properties of these functions and provide an 
evaluation of the corresponding amplitudes. 

\indent

\noindent
In terms of the effective couplings introduced in eq. \rf{eff_couplings},
these equations mean\footnote{We recall that these expressions hold in this form under the conditions 
listed after equation \rf{WIs}.}
\bea
C_{H\gamma\gamma}^{(u)} \!\!\!\!&=&\!\!\!\! - \frac{g e^2}{2} \frac{m_u}{M_W} \, e_u^2 \times 2 \left[
{\cal F}_{\gamma\gamma} (0 , 0 , M_H^2) + {\tilde{\cal F}}_{\gamma\gamma} (0 , 0 , M_H^2)
\right]\!   ,
\nonumber\\[0.2cm]
C_{H\gamma\gamma}^{(d)} = \frac{m_d}{m_s} \, C_{H\gamma\gamma}^{(s)} \!\!\!\!&=&\!\!\!\! 
- \frac{g e^2}{2} \frac{m_d}{M_W} \, e_d^2 \times 4 \left[
- {\cal F}_{\gamma\gamma} (0 , 0 , M_H^2) + 2 {\tilde{\cal F}}_{\gamma\gamma} (0 , 0 , M_H^2)
\right]\!   ,
\nonumber\\
\\
C_{H\gamma Z}^{(u)} \!\!\!\!&=&\!\!\!\! - \frac{g^2 e}{4} \frac{m_u}{c_{\rm w} M_W} \, e_u \times \frac{2}{3}\left[
- {\cal F}_{\gamma Z} (0 , M_Z^2 , M_H^2) + 2 (1 - 2 s_{\rm w}^2) {\cal F}_{\gamma\gamma} (0 , M_Z^2 , M_H^2) 
+ 2 (1 - 2 s_{\rm w}^2) {\tilde{\cal F}}_{\gamma\gamma} (0 , M_Z^2 , M_H^2)
\right]\!   ,
\nonumber\\[0.2cm]
C_{H\gamma Z}^{(d)} = \frac{m_d}{m_s} \, C_{H\gamma Z}^{(s)} \!\!\!\!&=&\!\!\!\! 
- \frac{g^2 e}{4} \frac{m_d}{c_{\rm w} M_W} \, e_d \times \frac{2}{3}\left[
- {\cal F}_{\gamma Z} (0 , M_Z^2 , M_H^2) + 2 (1 - 2 s_{\rm w}^2) {\cal F}_{\gamma\gamma} (0 , M_Z^2 , M_H^2) 
- 4 (1 - 2 s_{\rm w}^2) {\tilde{\cal F}}_{\gamma\gamma} (0 , M_Z^2 , M_H^2)
\right]\!    .
\nonumber
\eea
In view of these expressions, a remark is in order: the amplitudes for the $\gamma Z$ decay channel 
are not proportional to the coupling ${\rm v}_f$ of the $Z$ boson
to the fermions in the case of the light quarks. This is due to the fact that the current $J^\mu_{(Z)}$
decomposes into a flavour-singlet component and a flavour-nonsinglet one, with respective couplings to 
the quarks given by $-1/6$ ($1/6$ corresponds to the average electric charge within a quark doublet)
and $(1 - 2 s_{\rm w}^2) e_f$. The QCD dynamics makes a distinction between 
the flavour-singlet and flavour-nonsinglet channels. The proportionality to ${\rm v}_f$ occurs only
in situation where this difference, suppressed by the OZI rule \cite{OZI}, is absent. In perturbative QCD, this is the case 
at lowest and next-to-lowest \cite{Spira:1991tj,Gehrmann:2015dua,Bonciani:2015eua} orders, 
but it stops being true starting from order ${\cal O} (\alpha_s^2)$ onward (this holds also for the heavy quarks). 
As far as we are aware, perturbative QCD corrections for the $H\to\gamma Z$ amplitude beyond the order ${\cal O} (\alpha_s)$
have not been considered so far in the literature. 
In the non-perturbative treatment of the light quark flavours, this OZI-suppressed contribution is present from the start. 
Whether of perturbative or non-perturbative origin, it is usually considered to be small. In the non-perturbative regine, this feature finds 
a natural explanation in the limit where the number of colours $N_c$ becomes infinite \cite{tHooft:1973alw,Witten:1979kh},
where the OZI rule becomes exact. 
In the sequel, we will evaluate the non-perturbative contributions to the light-quark contributions in this limit. One may make the point 
about these OZI-suppressed contributions more visible upon rewriting the coefficients as (recall that
${\rm v}_s = {\rm v}_d$ and $e_s=e_d$)
\bea  
C_{H\gamma Z}^{(u)} \!\!\!\!&=&\!\!\!\!   
- \frac{g^2 e}{4 c_{\rm w}} \frac{m_u }{M_W} 
\times e_u
\bigg\{ 2 {\rm v}_u \left[ {\cal F}_{\gamma\gamma} (0 , M_Z^2 , M_H^2) + {\tilde{\cal F}}_{\gamma\gamma} (0 , M_Z^2 , M_H^2) \right]
\nonumber\\
&& \hspace{2.5cm}
+ \, \frac{1}{3} \left[ {\cal F}_{\gamma\gamma} (0 , M_Z^2 , M_H^2) + {\tilde{\cal F}}_{\gamma\gamma} (0 , M_Z^2 , M_H^2) 
- 2 {\cal F}_{\gamma Z} (0 , M_Z^2, M_H^2) \right]
\!\bigg\}   ,
\nonumber\\
\\
C_{H\gamma Z}^{(d)} = \frac{m_d}{m_s} \, C_{H\gamma Z}^{(s)} \!\!\!\!&=&\!\!\!\!   
- \frac{g^2 e}{4 c_{\rm w}} \frac{m_d}{M_W} 
\times e_d
\bigg\{ 2 {\rm v}_d \left[ - 2 {\cal F}_{\gamma\gamma} (0 , M_Z^2 , M_H^2) + 4 {\tilde{\cal F}}_{\gamma\gamma} (0 , M_Z^2 , M_H^2) \right]
\nonumber\\
&& \hspace{2.5cm}
- \, \frac{2}{3} \left[ {\cal F}_{\gamma\gamma} (0 , M_Z^2 , M_H^2) - 2 {\tilde{\cal F}}_{\gamma\gamma} (0 , M_Z^2 , M_H^2) 
+ {\cal F}_{\gamma Z} (0 , M_Z^2, M_H^2) \right]
\!\bigg\}   .
\nonumber
\eea
One clearly sees that proportionality to ${\rm v}_f$ is realized if and only if
\be
{\cal F}_{\gamma\gamma} (0 , M_Z^2 , M_H^2) = {\tilde{\cal F}}_{\gamma\gamma} (0 , M_Z^2 , M_H^2) = {\cal F}_{\gamma Z} (0 , M_Z^2, M_H^2)   .
\lbl{Zweig_condition}
\ee
Notice that if the condition \rf{Zweig_condition} is satisfied, the non-perturbative result for the effective coefficients
relative to the light quarks is obtained upon making the substitutions 
${\cal F}^{(f)} ( 0 , 0 , M_H^2) \longrightarrow {\cal F}_{\gamma\gamma} (0 , 0 , M_H^2)$,
${\cal F}^{(f)} ( 0 , M_Z^2 , M_H^2) \longrightarrow {\cal F}_{\gamma\gamma} (0 , M_Z^2 , M_H^2)$
in the corresponding perturbative one-loop expressions \rf{C_pert}.

\section{ }
\setcounter{equation}{0}

The general study of the decay amplitudes in the preceding section has left us with the task 
of evaluating the hadronic functions ${\cal F}_{\gamma\gamma} (q_1^2 , q_2^2, q_3^2)$, ${\tilde{\cal F}}_{\gamma\gamma} (q_1^2 , q_2^2, q_3^2)$ 
and ${\cal F}_{\gamma Z} (q_1^2 , q_2^2, q_3^2)$ in QCD with three flavours of massless quarks. In this chiral limit, these 
functions vanish exactly at each order of perturbative QCD. This follows from the fact that the correlators of the type $\langle S V V \rangle$,
where $S$ and $V$ denote a colour-singlet scalar density or vector current, respectively, are order parameters of 
the spontaneous breaking of the three-flavour chiral symmetry in QCD. Before turning to the evaluation of these functions
we will first show that they indeed do not vanish in the non-perturbative regime of QCD. In order to make full use 
of flavour symmetry, we introduce, in the chiral limit of QCD, the correlator

\bea 
\Gamma^{abc}_{\mu\nu} (q_1 , q_2) \!\!\!\!\!&=&\!\!\!\!
\int d^4 x \, e^{i q_1 \cdot x} \int d^4 y \, e^{i q_2 \cdot y}
\langle 0 \vert T \{ S^a (0) V^b_\mu (x) V^c_\nu (y) \} \vert 0 \rangle
\nonumber\\
\!\!\!\!\!&=&\!\!\!\!
d^{abc} \left[ {\cal F} (q_1^2 , q_2^2 , q_3^2) P_{\mu\nu} (q_1 , q_2) + 
{\cal G} (q_1^2 , q_2^2 , q_3^2) Q_{\mu\nu} (q_1 , q_2) \right]
\lbl{Gamma_abc}
\eea
defined in terms of the colour-singlet quark bilinears
\be
S^a = {\bar\psi} \frac{\lambda^a}{2} \psi , ~~~ 
V_\mu^a = {\bar\psi} \frac{\lambda^a}{2} \gamma_\mu \psi,  
\lbl{Sa_and_Va}
\ee
where $\psi$ is defined in eq. \rf{psi_M_Q} and $\lambda^a$, $a=1,2,\ldots,8$ denote the Gell-Mann matrices acting 
in flavour space, with 
\be
d^{abc} =  \frac{1}{4} \, {\rm tr} ( \lambda^a \{ \lambda^b , \lambda^c \} )   .
\lbl{d^abc}
\ee
A full calculation of this correlator is unfortunately beyond reach of our present knowledge of the non-perturbatice regime of QCD.
For the time being, let us point out an important feature of this three-point function: the absence of perturbative contributions 
implies that the correlator \rf{Gamma_abc} has a smooth behaviour at short distances. Indeed, from the operator-product 
expansion, one obtains, when the momenta $q_1$, $q_2$ and $q_3$ are all in the deep-Euclidean region, that
\cite{Moussallam:1994at,Kadavy:2020hox}
\be
\lim_{\lambda\to\infty} \lambda^2 \Gamma^{abc}_{\mu\nu} (\lambda q_1 , \lambda q_2) =
- \, d^{abc} \, \frac{\langle {\bar q} q \rangle_0}{2 q_1^2 q_2^2 q_3^2} \Big\{
\Big[ (q_1^2 + q_2^2) [1 + {\cal O} (\alpha_s) ] + q_3^2 [1 + {\cal O} (\alpha_s) ] \Big] P_{\mu\nu} (q_1 , q_2) 
+ 2 [1 + {\cal O} (\alpha_s) ] Q_{\mu\nu} (q_1 , q_2) \Big\}
+  \cdots   ,
\lbl{SD}
\ee
where the ellipsis stands for subleading contributions to the short-distance behaviour and $\langle {\bar q} q \rangle_{\! _0}$ 
denotes the single-flavour quark condensate in the chiral limit,
\be
\langle {\bar q} q \rangle_{\! _0} \equiv \langle {\bar u} u \rangle_{\! _0} = \langle {\bar d} d \rangle_{\! _0} = \langle {\bar s} s \rangle_{\! _0}  .
\ee 
The presence of the quark condensate as a multiplicative factor of the leading term in the short-distance expansion 
reflects the fact that the correlator $\Gamma^{abc}_{\mu\nu} (q_1 , q_2)$ is an order parameter of the spontaneous breaking 
of the three-flavour chiral symmetry.
The ${\cal O} (\alpha_s)$ perturbative corrections to the Wilson coefficients are not particularly important for our purposes: 
since the leading operator, the quark condensate, carries the same anomalous dimension as the scalar density $S^a$, the corresponding 
Wilson coefficients have no anomalous dimension \cite{Moussallam:1994at}. Likewise, in the limit where only
the momenta $q_1$ and $q_3$ belong to the deep-Euclidean region, one obtains
\be
\lim_{\lambda\to\infty} \lambda \Gamma^{abc}_{\mu\nu} ( q_1 , \lambda q_2) =
d^{abc} \, \Pi_{VT} (q_1^2) \frac{1}{q_2^2} [1 + {\cal O} (\alpha_s) ] P_{\mu\nu} (q_1 , q_2) + \cdots ,
\lbl{SD2}
\ee
where the vector-tensor correlator $\Pi_{VT} (q^2)$ in the chiral limit, defined as
\be
\frac{i}{2} \int d^4 x \, e^{i q \cdot x} \langle 0 \vert T \Big\{ V^a_\mu (x)  
\Big( {\bar\psi} \frac{\lambda^b}{2} [ \gamma_\nu , \gamma_\rho ] \psi \Big) (0) \Big\} \vert 0 \rangle =
\delta^{ab} \Pi_{VT} (q^2) ( q_\nu \eta_{\mu\rho} - q_\rho \eta_{\mu\nu} )  ,
\ee
is also an order parameter  of the spontaneous chiral-symmetry breaking. Moreover, $\Pi_{VT} (0)$ is given 
by the product of the quark condensate  multiplied by the magnetic
susceptibility of the quarks \cite{Ioffe:1983ju}. The latter has been determined, for instance, 
by lattice QCD \cite{Bali:2012jv} or QCD sum rules \cite{Ball:2002ps}, and definitely does not vanish.

\indent

\noindent
The results \rf{SD} and \rf{SD2} thus tell us at once and without ambiguity that the pure octet
function ${\cal F}_{\gamma\gamma}$ in equation \rf{vertex_def} does not vanish identically in QCD. 
Let us stress that this statement holds at the non-perturbative level and does not require 
to know the correlator \rf{Gamma_abc} in the full ranges of the momenta it depends upon.
A similar argument, leading to the same conclusion, applies to the remaining  two functions 
introduced in eq. \rf{vertex_def}, ${\tilde{\cal F}}_{\gamma\gamma}$ and 
${\cal F}_{\gamma Z}$, thus validating the claim made after eq. \rf{amps_uds}:
not only does the contribution from the light quarks to the amplitudes for $H\to\gamma\gamma$
or $H \to \gamma Z$ not vanish, but it starts with terms that are \textit{linear} in the 
quark masses, and not quadratic.

\section{ }
\setcounter{equation}{0}

As already alluded to, evaluating these three-point functions beyond their asymptotic short-distance regimes \rf{SD} and \rf{SD2}
turns into an arduous task. Some insight into this issue can possibly be gained if one considers the limit where the number of colours $N_c$ becomes 
infinite as an approximation to the case $N_c=3$ relevant for QCD \cite{tHooft:1973alw,Witten:1979kh}. 
A first useful feature of the large-$N_c$ limit is that the Gell-Mann--Okubo octet symmetry is actually enlarged to a 
nonet symmetry: the definitions \rf{Sa_and_Va} and \rf{d^abc} can be extended upon introducing the matrix $\lambda^0 = \sqrt{2/3} \, 1\!\!1$,
where $1\!\!1$ denotes the unit $3\times 3$ matrix, so that the definition \rf{Gamma_abc} includes the correlators 
involving also the singlet operators $S^0$ and $V^0$, with $d^{\,0ab}=\sqrt{2/3} \, \delta^{ab}$.
Making now contact with the vertex functions defined in eq. \rf{vertex_def}, we find that
\be
{\cal F} (q_1^2 , q_2^2 , q_3^2) =
{\cal F}_{\gamma \gamma} (q_1^2 , q_2^2 , q_3^2) =
{\cal F}_{\gamma Z} (q_1^2 , q_2^2 , q_3^2) =
{\tilde{\cal F}}_{\gamma\gamma} (q_1^2 , q_2^2 , q_3^2)
\ee
in the combined large-$N_c$ and chiral limits (identical equalities also hold between the functions ${\cal G}_{\gamma \gamma}$,
${\cal G}_{\gamma Z}$, ${\tilde{\cal G}}_{\gamma\gamma}$ and ${\cal G}$).

\indent

\noindent 
The general structure of the three-point function \rf{Gamma_abc} becomes rather simple in the large-$N_c$ limit:
it is given by an infinite number of poles due to zero-width bound states in each channel \cite{Witten:1979kh}. The quantum numbers 
of these bound states are determined by those of the operators $V^a_\mu$ and $S^a$, i.e. $J^{P}=1^{-}$ states in  
each of the vector channels, and $J^{P}=0^{+}$ states in the scalar channel.
Usually, one makes the approximation that the lowest-mass resonance in each channel plays the most 
important role, given that in practice the asymptotic behaviour \rf{SD} is expected to set in quickly when, 
say, $-q_i^2 \gapprox 2~{\rm GeV}^2$, $i=1,2,3$. This ``lowest-meson dominance'' (LMD) approximation to 
large-$N_c$ QCD \cite{Peris:1998nj,Golterman:1999au,Knecht:1999gb} leads to the quite simple expression \cite{Moussallam:1994at}
\be
{\cal F}_{\rm LMD} (q_1^2 , q_2^2 , q_3^2) =
\frac{a + b \, (q_1^2 + q_2^2) + c \, q_3^2}{(q_1^2 - M_V^2) (q_2^2 - M_V^2) (q_3^2 - M_S^2)}   ,
\ee
involving only the three so-far unknown parameters $a,b,c$.
The matching with the short-distance behaviour \rf{SD} allows to fix two of them,
\be
b = c = - \frac{1}{2} \, \langle {\bar q} q \rangle_{\! _0}   .
\ee
Finally, as pointed out in ref. \cite{Moussallam:1994at}, the last parameter is fixed by a sum rule \cite{Knecht:1994ug}
involving the $e^+ e^-$ partial widths of the vector resonances (${\hat m} \equiv (m_u + m_d)/2$),
\be
a = - \frac{9}{5} M_V^4 M_S^2 \frac{\tilde c}{m_s - {\hat m}} , ~~~ 
{\tilde c} =  \frac{5}{16 \pi \alpha^2} \left[ \frac{\Gamma_{\rho\to e^+ e^-}}{M_\rho} 
- 3 \frac{\Gamma_{\omega\to e^+ e^-}}{M_\omega} - 3 \frac{\Gamma_{\phi\to e^+ e^-}}{M_\phi}\right]   . 
\ee
This gives, within this LMD approximation to the large-$N_c$ limit of QCD, the expression
\be
{\cal F}_{\rm LMD} (q_1^2 , q_2^2 , q_3^2) = - \frac{1}{10} \, \frac{1}{m_s - {\hat m}} \,
\frac{18 M_V^4 M_S^2 {\tilde c} + 5 (m_s - {\hat m}) \langle {\bar q} q \rangle_{\! _0} (q_1^2 + q_2^2 + q_3^2)}{(q_1^2 - M_V^2) (q_2^2 - M_V^2) (q_3^2 - M_S^2)} .
\lbl{calF_LMD}
\ee
Interestingly enough, this simple expression, constructed to satisfy the short-distance constraint 
\rf{SD}, actually happens to satisfy \rf{SD2} as well, provided one uses the same LMD approximation
for the vector-tensor correlator, i.e. \cite{Knecht:2001xc}
\be
\Pi_{VT}^{\rm LMD} (q^2) = - \frac{\langle {\bar q} q \rangle_0}{q^2 - M_V^2}   .
\ee

\indent

\noindent
The expressions of the effective $H-\gamma-\gamma$ and $H-\gamma - Z$ couplings obtained within this framework
thus read
\bea\lbl{couplings_LMD}
C_{H\gamma\gamma}^{(u)} \!\!\!\!&=&\!\!\!\! 
- \frac{g e^2}{2} \frac{m_u}{M_W} \times 4 e_u^2  {\cal F}_{\rm LMD} ( 0 , 0 , M_H^2)     ,
\nonumber\\[0.2cm]
C_{H\gamma\gamma}^{(d)} = \frac{m_d}{m_s} \, C_{H\gamma\gamma}^{(s)} \!\!\!\!&=&\!\!\!\! 
- \frac{g e^2}{2} \frac{m_d}{M_W} \times 4 e_d^2 {\cal F}_{\rm LMD} (0 , 0 , M_H^2)  ,
\nonumber\\
\\
C_{H\gamma Z}^{(u)} \!\!\!\!&=&\!\!\!\! 
 - \frac{g^2 e}{4 c_{\rm w}} \frac{m_u}{M_W} \times 4 e_u {\rm v}_u
{\cal F}_{\rm LMD} (0 , M_Z^2 , M_H^2)    ,
\nonumber\\[0.2cm]
C_{H\gamma Z}^{(d)} = \frac{m_d}{m_s} \, C_{H\gamma Z}^{(s)} \!\!\!\!&=&\!\!\!\! 
 - \frac{g^2 e}{4 c_{\rm w}} \frac{m_d}{M_W} \times 4 e_d {\rm v}_d
{\cal F}_{\rm LMD} (0 , M_Z^2 , M_H^2)     .
\nonumber
\eea
The important point to notice here is that the product of ${\cal F}_{\rm LMD} (q_1^2 , q_2^2 , q_3^2)$ with a quark mass
gives a quantity that does no longer depend on any renormalization scale. The expressions above of the non-perturbative coefficients
$C_{H\gamma\gamma}^{(q)}$ and $C_{H\gamma Z}^{(q)}$ corresponding to the light quarks $q=u,d,s$ are thus free of any renormalization 
ambiguity, since they involve only the ratio of quark masses or the product of a quark mass with the single-flavour quark condensate,
quantities which are both invariant under the renormalization group.

\section{ }
\setcounter{equation}{0}

We now turn to the numerical evaluation of the light-quark contribution
to the effective $H - \gamma - \gamma$ and $H - \gamma - Z$ couplings within
the theoretical framework outlined in the preceding section, i.e. using the 
expressions \rf{calF_LMD} and \rf{couplings_LMD}.
For the resonance masses, we take $M_V = 770~{\rm MeV}$ for the lowest vector state 
and $M_S=1400~{\rm MeV}$ for the lowest scalar state.\footnote{Ref. \cite{ParticleDataGroup:2024cfk}
lists several lighter scalar states, but it has been argued \cite{Oller:1998zr,Cirigliano:2003yq,Bijnens:2014lea}
that these states are generated dynamically at next-to-leading orders in the large-$N_c$ expansion, and therefore 
disappear from the spectrum in the limit $N_c \to \infty$. In practice, the numerical value one takes for 
$M_S$ does not play a crucial role, as long as it is of order $\sim 1~{\rm GeV}$.} 
For the value of ${\tilde c}$ we take the value given in \cite{Knecht:2018sci}\footnote{With the 
values of the decay widths given in the more recent issue of the Review of Particle Physics \cite{ParticleDataGroup:2024cfk},
the central value of ${\tilde c}$ is slightly increased, \textit{viz.} $0.52$ instead of $0.46$. Again, this has 
no effect on the values displayed in table \ref{tab:results}. See also the discussion below.}
\be
{\tilde c} = 4.6(0.8) \cdot 10^{-3}   .
\ee
For the products of a quark mass times the condensate we use \cite{FLAG:2021npn}
\be
m_u \langle {\bar q} q \rangle_{\! _0} \sim m_d \langle {\bar q} q \rangle_{\! _0} = -(0.091 ~ {\rm GeV})^4 ~~~
m_s \langle {\bar q} q \rangle_{\! _0} = -(0.208 ~ {\rm GeV})^4   ,
\ee
based on the results from lattice QCD of refs. \cite{MILC:2009ltw,McNeile:2010ji,BMW:2010skj,BMW:2010ucx,Bazavov:2010yq,RBC:2014ntl,Bruno:2019vup}.
The values of the effective coefficients $C_{\gamma\gamma}^{(q)}$ and $C_{\gamma Z}^{(q)}$ for 
the light quarks $q=u,d,s$ obtained from 
the expressions \rf{couplings_LMD} and these input values are given in the left-hand part of table \ref{tab:results},
together with the results that follow from the lowest-order perturbative expressions \rf{C_pert} and \rf{calF_pert}.
The latter were obtained with the following values of the quark masses \cite{ParticleDataGroup:2024cfk,FLAG:2021npn}
(${\overline m}(\nu)$ denotes the runnung mass in the ${\overline{\rm MS}}$ scheme, for the top 
quark we take the pole mass):
\bea
&&\hspace{2.cm}
m_t = 172.57~{\rm GeV} , ~~~ {\overline m}_b ({\overline m}_b) = 4.18~{\rm GeV} , ~~~ {\overline m}_c ({\overline m}_c) = 1.28~{\rm GeV} 
\nonumber\\
&&
{\overline m}_s (\nu = 2~{\rm GeV}) = 92.74~{\rm MeV} , ~~~ {\overline m}_d (\nu = 2~{\rm GeV}) = 4.69~{\rm MeV} , ~~~ 
{\overline m}_u (\nu = 2~{\rm GeV}) = 2.20~{\rm MeV}.
\eea
\begin{table}[t]
\begin{center}
{\small
\begin{tabular}{c||c|c}
\hline
   &  pQCD &  LMD
\\
\hline
& & 
\\[-0.2cm]
$ C_{H\gamma\gamma}^{(t)} $  &  $ -8.66 \cdot 10^{-6} $  &  $ - $
\\[0.15cm]
$ C_{H\gamma\gamma}^{(b)} $  &  $ (1.13 - 1.49 i) \cdot 10^{-7} $  &  $ - $
\\[0.15cm]
$ C_{H\gamma\gamma}^{(c)} $  &  $ (9.22 - 7.57 i) \cdot 10^{-8} $  &  $ - $
\\[0.15cm]
$ C_{H\gamma\gamma}^{(s)} $  &  $ (3.35 - 1.56 i) \cdot 10^{-10} $  &  $ -4.41 \cdot 10^{-7} $
\\[0.15cm]
$ C_{H\gamma\gamma}^{(d)} $  &  $ (1.77 - 0.57 i) \cdot 10^{-12} $  &  $ -2.23 \cdot 10^{-8} $
\\[0.15cm]
$ C_{H\gamma\gamma}^{(u)} $  &  $ (1.81 - 0.53 i) \cdot 10^{-12} $  &  $ -4.18 \cdot 10^{-8} $
\\[0.15cm]
\hline
& & 
\\[-0.2cm]
$ C_{H\gamma Z}^{(t)} $  &  $ -3.30 \cdot 10^{-6} $  &  $ - $
\\[0.2cm]
$ C_{H\gamma Z}^{(b)} $  &  $ (6.82 - 3.83 i) \cdot 10^{-8} $  &  $ - $
\\[0.15cm]
$ C_{H\gamma Z}^{(c)} $  &  $ (1.05 - 0.41 i) \cdot 10^{-8} $  &  $ - $
\\[0.15cm]
$ C_{H\gamma Z}^{(s)} $  &  $ (7.93 - 1.89 i) \cdot 10^{-11} $  &  $ 6.20 \cdot 10^{-11} $
\\[0.15cm]
$ C_{H\gamma Z}^{(d)} $  &  $ (2.94 - 0.48 i) \cdot 10^{-13} $  &  $ 3.14 \cdot 10^{-12} $
\\[0.15cm]
$ C_{H\gamma Z}^{(u)} $  &  $ (8.06 - 1.22 i) \cdot 10^{-14} $  &  $ 1.70 \cdot 10^{-12} $
\end{tabular}
\hspace{2.5cm}
\begin{tabular}{c||c|c}
\hline
   &  pQCD &  LMD 
\\
\hline
& & 
\\[-0.2cm]
$ h_{1;t}^{\gamma\gamma} $  &  $ -1.04 \cdot 10^{-1} $  &   $-$  
\\[0.2cm]
$ h_{1;b}^{\gamma\gamma} $  &  $ (1.35 - 1.79 i) \cdot 10^{-3} $  &   $-$  
\\[0.2cm]
$ h_{1;c}^{\gamma\gamma} $  &  $ (1.11 - 0.91 i) \cdot 10^{-3} $  &   $-$  
\\[0.2cm]
$ h_{1;s}^{\gamma\gamma} $  &  $ (4.02 - 1.88 i) \cdot 10^{-6} $  &  $-5.29 \cdot 10^{-3}$  
\\[0.2cm]
$ h_{1;d}^{\gamma\gamma} $  &  $ (2.13 - 0.68 i) \cdot 10^{-8} $  &  $-2.68 \cdot 10^{-4}$  
\\[0.2cm]
$ h_{1;u}^{\gamma\gamma} $  &  $ (2.17 - 0.64 i) \cdot 10^{-8} $  &  $-5.02 \cdot 10^{-4}$  
\\[0.2cm]
\hline
& & 
\\[-0.2cm]
$ h_{1;t}^{\gamma Z} $  &  $ 1.86 \cdot 10^{-2} $  &  $-$  
\\[0.2cm]
$ h_{1;b}^{\gamma Z} $  &  $ (-3.84 + 2.16 i) \cdot 10^{-4} $  &  $-$  
\\[0.2cm]
$ h_{1;c}^{\gamma Z} $  &  $ (-5.91 + 2.33 i) \cdot 10^{-5} $  &  $-$  
\\[0.2cm]
$ h_{1;s}^{\gamma Z} $  &  $ (-4.47 + 1.06 i) \cdot 10^{-7} $  &  $-3.49 \cdot 10^{-7}$  
\\[0.2cm]
$ h_{1;d}^{\gamma Z} $  &  $ (-1.66 + 0.27 i) \cdot 10^{-9} $  &  $-1.77 \cdot 10^{-8}$   
\\[0.15cm]
$ h_{1;u}^{\gamma Z} $  &  $ (-4.54 + 0.69 i) \cdot 10^{-10} $  &  $-9.55 \cdot 10^{-9}$   
\\[0.05cm]
\end{tabular}
}
\end{center}
\caption{Numerical results for the effective coefficients $C_{H\gamma X}^{(f)}$
(in ${\rm GeV}^{-1}$, table on the left) and $h_{1;f}^{\gamma X}$ (in GeV, table on the right)  
for $X=\gamma,Z$ with the input values given in the text. The columns labelled `pQCD' correspond 
to the values at lowest order of perturbation theory, whereas the columns labelled `LMD'
show the values obtained for the light quarks with the non-perturbative form factor 
${\cal F}_{\rm LMD}$ of eq. \rf{calF_LMD}.}
\label{tab:results}
\end{table}
Some authors \cite{Hernandez-Juarez:2025ees,Hernandez-Juarez:2024iwe} express the amplitudes in terms of coefficients $h_{1;f}^{\gamma\gamma}$ and 
$h_{1;f}^{\gamma Z}$ given by
\be
h_{1;f}^{\gamma X} = \epsilon_X \, \frac{M_H^2 - M_X^2}{2 g} \, C_{H\gamma X}^{(f)} , ~~~ X=\gamma , Z , ~~~ \epsilon_\gamma = +1, ~ \epsilon_Z = -1.
\ee
To allow for a straightforward comparison with these references we show the results we obtain for 
these coefficients in the right-hand part of table \ref{tab:results}.
We now list a few comments concerning the results displayed in this table:

\begin{itemize}
\item If one instead uses the same input values
\be
m_u = m_d = 8~{\rm MeV} , ~~~ m_s = 150~{\rm MeV}
\ee
for the light-quark masses as in \cite{Hernandez-Juarez:2025ees}, one obtains pQCD results
that agree with those quoted in this reference, i.e.
\be
h_{1;s}^{\gamma\gamma} = (9.06 - 4.58 i) \cdot 10^{-6} ~ {\rm GeV}, ~~~
h_{1;d}^{\gamma\gamma} = (5.54 - 1.87 i) \cdot 10^{-8} ~ {\rm GeV}, ~~~ 
h_{1;u}^{\gamma\gamma} = (2.22 - 0.75 i) \cdot 10^{-7} ~ {\rm GeV}, 
\ee 
and
\be
h_{1;s}^{\gamma Z} = (-1.08 + 0.28 i) \cdot 10^{-6} ~ {\rm GeV}, ~~~
h_{1;d}^{\gamma Z} = (-4.56 + 0.79 i) \cdot 10^{-9} ~ {\rm GeV}, ~~~
h_{1;u}^{\gamma Z} = (-5.25 + 0.91 i) \cdot 10^{-9} ~ {\rm GeV}.
\ee

\item In the case of $H \to \gamma \gamma$, one notices an increase 
of the absolute value of the sum  $C_{H\gamma\gamma}^{(u)}+C_{H\gamma\gamma}^{(d)}+C_{H\gamma\gamma}^{(s)}$
by at least three orders of magnitude as compared to the lowest-order perturbative evaluation. 
In the case of $H \to \gamma Z$ no such dramatic changes 
are present. 

\item From a numerical point of view, the numerator of ${\cal F}_{\rm LMD} (0,q^2,M_H^2)$ is dominated 
by the second term, proportional to $(q^2 + M_H^2)$:  for the values $q^2=0$ or $q^2=M_Z^2$ of interest here, 
its contribution is larger by about three to four orders of magnitude than the one proportional to ${\tilde c}$. 
Therefore, one has
\be
{\cal F}_{\rm LMD} (0,M_Z^2,M_H^2) \sim \frac{1 + \frac{M_Z^2}{M_H^2}}{1 - \frac{M_Z^2}{M_V^2}} \times {\cal F}_{\rm LMD} (0,0,M_H^2)
\sim - 10^{-4} \times {\cal F}_{\rm LMD} (0,0,M_H^2)   ,
\ee
with ${\cal F}_{\rm LMD} (0,0,M_H^2) \sim 0.03$. This explains why there is no sizeable enhancement
in the effective coefficients for $H \to \gamma Z$ as compared to their perturbative values.

\item
The comparison of the predictions (without QCD corrections for the perturbative contributions) for the decay rates gives
\be
\Gamma (H \to \gamma \gamma)_{\rm LMD} = 0.97 \cdot \Gamma (H \to \gamma \gamma)_{\rm pQCD},
\lbl{rates_comp}
\ee
whereas $\Gamma (H \to \gamma Z)$ is not affected.

\item The effective coefficients for the light quarks are purely real in the description of non-perturbative QCD effects 
adopted here. This is because in the large-$N_c$ limit the only singularities that survive are poles.
In order to generate imaginary parts, one would also need to account for multi-particle intermediate states. 
These would introduce imaginary parts through the cut singularities they produce in the $\langle SVV \rangle$
three-point function. This, however, requires to go beyond the strict large-$N_c$ limit and is beyond the 
scope of the present study.

\end{itemize}

\section{ }
\setcounter{equation}{0}

To summarize, we have shown that, contrary to the claim made in ref. \cite{Hernandez-Juarez:2025ees},
the contribution from the light quarks to the decay amplitudes for $H \to \gamma \gamma$ and 
$H \to \gamma Z$ does not vanish at all in the standard model. Moreover, we have shown that 
the leading behaviour of this contribution is only linear in the quark masses instead of quadratic, 
the behaviour exhibited by a naive perturbative evaluation of the amplitudes. These statements are 
exact since they rest only on the identification of the $\langle S V V \rangle$ correlators like \rf{Gamma_abc} 
as the appropriate QCD Green's functions to describe the contribution from the light quarks to these decay modes, 
and on their short-distance properties in QCD.
The claim made by the authors of ref. \cite{Hernandez-Juarez:2025ees} follows from
considering contributions (mediating the non-perturbative $H - \gamma - X$ coupling, $X=\gamma,Z$,
through the exchange of pseudoscalar states) that would rather belong to the $\langle P V V \rangle$ three-point 
function, where $P={\bar \psi} i \gamma_5 \psi$ is the pseudoscalar density. Obviously, the 
standard-model Brout-Englert-Higgs scalar boson couples to two photons or a photon and a $Z$ 
via the $\langle S V V \rangle$ three-point function.\footnote{The $\langle P V V \rangle$
three-point function would become relevant in cases, like two-Higgs doublet models, where
the scalar sector contains also a $J^{P}=0^{-}$ state. For a treatment of the $\langle P V V \rangle$
three-point functions along similar lines, see refs. \cite{Knecht:1999gb,Knecht:2001xc,Kadavy:2020hox}.}

\indent 

\noindent 
We have evaluated this three-point function in the LMD
approximation to the large-$N_c$ limit of QCD.
In the two-photon channel, this results in quite a substantial enhancement, by several
orders of magnitude, of the contribution from the light quarks.
A handwaving argument as to why some enhancement is to be expected can be given as follows: in the 
perturbative treatment, the second power of the quark mass $m_q$ is produced by the loop integral;
in the non-perturbative treatment, one expects that the scale that emerges is a typical hadronic
scale $\Lambda_{\rm had} \sim 1~{\rm GeV}$, so that $m_q^2$ becomes $m_q \Lambda_{\rm had}$. 
Within the large-$N_c$/LMD representation of the relevant three-point function considered here,
the enhancement we find in the case of $H \to \gamma \gamma$  goes well beyond this naive 
expectation, and is actually driven by the presence of another scale, $M_H$. In
the case of $H \to \gamma Z$ a similar large enhancement
is thwarted by the presence of an additional large scale, the mass $M_Z$.
While the quantitative conclusions concerning the decay rates following from this approximate description of the three-point 
functions should certainly be taken \textit{cum grano salis}, a deeper understanding of how these different scales, 
a low scale $\Lambda_{\rm had}$ related to the virtual 
quark and gluon degrees of freedom an the one hand, and a high scale like $M_H$ or $M_Z$ represented by the external 
states on the other hand, act upon each other at a non-perturbative level is necessary, and represents a real challenge.
Furthermore, another by-product of the large-$N_c$ representation of the form factors in terms of simple poles
is the absence of imaginary parts in the light-quark contribution to the decay amplitudes. To generate imaginary parts 
requires to take also multi-particle intermediate states into account, a sub-leading effect in the $1/N_c$ expansion 
and a task that goes well beyond the scope of this Letter.

\indent 

\noindent
The main conclusion of this Letter, besides clarifying the issue concerning the contribution of the light quarks
to the $H \to \gamma\gamma$ and $H \to \gamma Z$ decay amplitudes in the standard model, is therefore also
to point out that in the case of $H\to\gamma\gamma$ this contribution can potentially be much larger than expected so far. 
This observation may become important if in the future the precision on the branching 
ratio for $H \to \gamma \gamma$ approaches the percent \cite{ATLAS:2025-014},\footnote{We thank M. Talby
for pointing out this reference to us.} or even 
the sub-percent \cite{FCC:2025,LinearColliderVision:2025hlt}, level.
The contribution from the light quarks would then definitely need to be reconsidered under close scrutiny.

\end{document}